\documentclass[twocolumn,twocolappendix]{aastex631}

\usepackage{graphicx} 
\usepackage{amsmath}
\usepackage{subfigure}
\usepackage{amssymb}

\newcommand{\bn}{\vec{b}_n}
\newcommand{\Vn}{V_n}
\newcommand{\krisp}{\texttt{KRISP}}

\begin{document}
\title{Kernel Methods for Interferometric Imaging}
\date{\today}

\author[0000-0003-1035-3240]{Dimitrios Psaltis}
\affiliation{School of Physics, Georgia Institute of Technology, 837 State St NW, Atlanta, GA 30332, USA}

\author[0000-0003-4413-1523]{Feryal \"Ozel}
\affiliation{School of Physics, Georgia Institute of Technology, 837 State St NW, Atlanta, GA 30332, USA}

\author[0000-0003-0387-6617]{Yassine Ben Zineb}
\affiliation{School of Physics, Georgia Institute of Technology, 837 State St NW, Atlanta, GA 30332, USA}

\begin{abstract}
Increasing the angular resolution of an interferometric array requires placing its elements at large separations. This often leads to sparse coverage and introduces challenges to reconstructing images from interferometric data. We introduce a new interferometric imaging algorithm, \krisp, that is based on Kernel methods, is statistically robust, and is agnostic to the underlying image. The algorithm reconstructs the complete Fourier map up to the maximum observed baseline length based entirely on the data without tuning by a user or training on prior images and reproduces images with high fidelity. \krisp\ works efficiently for many sparse array configurations even in the presence of significant image structure as long as the typical baseline separation is comparable to or less than the correlation length of the Fourier map, which is inversely proportional to the size of the target image. 
\end{abstract}

\section{Introduction}

Interferometric imaging is a powerful technique that is frequently used in astronomy to achieve high angular resolution \citep{TMS2017}. Currently, there are multiple Earth- and space-based arrays in a variety of wavebands that reach $\sim 10$'s of $\mu$as resolution \citep{Kardashev2013,EHT_M87_II,Eisenhauer2023}. 

All these interferometric arrays measure complex visibilities between pairs of telescopes, which are equal to the complex Fourier components of the images at spatial frequencies determined by the baseline lengths between the telescopes and their relative orientations. The fidelity of the images reconstructed from such measurements depends on the extent and density of coverage of the Fourier space (or $u-v$ plane) by the array elements, which every array design aims to maximize. However, there is often a trade-off between sparseness and resolution of an array: reaching the highest angular resolution requires increasing the separation between the array elements, but that often leads to substantial gaps in the Fourier plane. This is especially true for global arrays, such as the Event Horizon Telescope \citep{EHT_M87_II}, or space-based interferometers, such as RadioAstron \citep{Kardashev2013}, where geographical or orbital constraints lead to sparse coverage. 

A variety of methods have been developed to handle image reconstruction for sparse interferometers. 
Early methods, such as CLEAN \citep{Hogbom1974,Clark1980} and Regularized Maximum Likelihood (e.g., \citealt{Baron2010,Chael2016,Akiyama2017}), use building blocks in the image domain to construct sky brightness distributions that are in agreement with the observed visibilities. More recently, machine-learning imaging approaches that utilize convolutional neural networks~\cite[CNNs;][]{Schmidt2022,Connor2022}, generative adversarial networks~\cite[GANs;][]{Geyer2023}, or deep learning~\citep{Lai2024}, have also been explored. In contrast, an alternative approach introduced in a recent algorithm, \texttt{PRIMO}, learns the building blocks of possible Fourier maps of images via training from simulations to reconstruct the full Fourier domain up to the maximum observed baseline length \citep{Medeiros2023a}. The latter approach works because, for any compact source, the finite size of the image introduces a unique correlation length in the Fourier domain and, as long as the gaps in the Fourier domain are smaller than the correlation length, a high fidelity image can be reconstructed \citep{Psaltis2024}. 

This fact allows us to go a step further in detaching the imaging algorithm from any particular training set and instead use kernel regression to effectively reconstruct the Fourier maps between the observed data points \citep{Rasmussen2006}. Kernel methods have the combined advantage of fitting the data with an infinite and complete set of functions and, at the same time, depend only on a small number of hyperparameters to establish the set of functions and make predictions within a correlation length. 

In this paper, we introduce a new algorithm called Kernel Regression Imaging for Sparse Patterns, or \krisp. Because the method is based on Kernel regression, it is agnostic to the target image and is based on a statistically robust foundation. We show that \krisp\ recovers the complete Fourier map up to the maximum baseline length of observations even with sparse sampling without the need of regularizers or user adjustments. We focus as our examples on the reconstruction of horizon-scale black hole images, for which the array size and sparseness introduce particularly acute challenges. 

In \S2, we review the basic properties of Kernel methods. In \S3, we introduce the \krisp\ algorithm and in \S4, we introduce several examples of sparse sampling configurations motivated by Earth- and space-based interferometers as well as test images. In \S5, we explore the impact of sparseness on the fidelity of the reconstruction and, in \S6, the effect of data uncertainties and of significant image complexity. 

\section{2D Kernel Methods}

Our goal is to fill in the gaps between data points in the $u-v$ plane with an algorithm that imposes a minimal number of assumptions. The data points are unevenly spaced and result from the sparse sampling provided by the observational configuration. Kernel methods provide a robust but flexible approach for achieving this goal.  

Let’s consider $N$ data points at baseline locations $\bn=(u,v)_n$ and complex visibilities $\Vn$, with $n=1, ..., N$. In principle, we can interpolate between the data points using a linear combination of K functions $\phi_i(\vec{b})$ such that 
\begin{equation}
V(\vec{b})=\sum_{i=1}^K w_i \phi_i(\vec{b})    
\end{equation}
and minimize a log-likelihood, such as a $\chi^2$
statistic
\begin{equation}
    \chi^2=\sum_{n=1}^N \left[\sum_{i=1}^K w_i \phi_i(\bn)-\Vn\right]^2+\lambda_r \sum_{i=1}^K w_i^2.
\end{equation}
The last term introduces a potential regularizer with weight $\lambda_r$ that is necessary when $K>N$. 

Because we are not interested in the fit parameters themselves but only on the value of the function at some location $\vec{b}$ between the data points, it is possible to show that the result depends only on the scalar product between the functions \citep{Bishop2006}
\begin{equation}
k(\vec{b},\vec{b}^\prime)=\sum_{i=1}^K \phi_i(\vec{b})\phi_i(\vec{b}^\prime)
\label{eq:kernel}
\end{equation}
referred to as the kernel. We can then write 
\begin{equation}
V(\vec{b})=\sum_{n,m} k(\vec{b},\vec{b}_n) \left[K_{nm}+\lambda_r\delta_{nm}\right]^{-1} V_m\;,
\label{eq:prediction_noerror}
\end{equation}
where $K_{nm}\equiv k(\vec{b}_n,\vec{b}_m)$ is the Gram matrix that measures the correlation between pairs of data points. 

Equation~(\ref{eq:prediction_noerror}) ignores uncertainties both in the measurements and in the model predictions. We are interested in generalizing this equation to incorporate both of these uncertainties. Assuming that the uncertainties in the measurements are described by Gaussian functions with the same standard deviation, i.e., the mean value of the $n-$th visibility is $\Vn$ and its standard deviation is $\sigma$, the posterior distribution for the prediction at an interpolated location $\vec{b}$ is also a Gaussian, with a mean given by 
\begin{equation}
<V(\vec{b})>=k^T K_{nm}^{-1} V_m
\label{eq:Vpred}
\end{equation}
and standard deviation
\begin{equation}
\sigma_b^2=\sigma^2-k^T K_N^{-1} k\;.
\label{eq:Verror}
\end{equation}
In this expression, we set the regularizer to zero for reasons that we will discuss below. 

\begin{figure*}[t]
\centering
\includegraphics[width=0.32\textwidth]{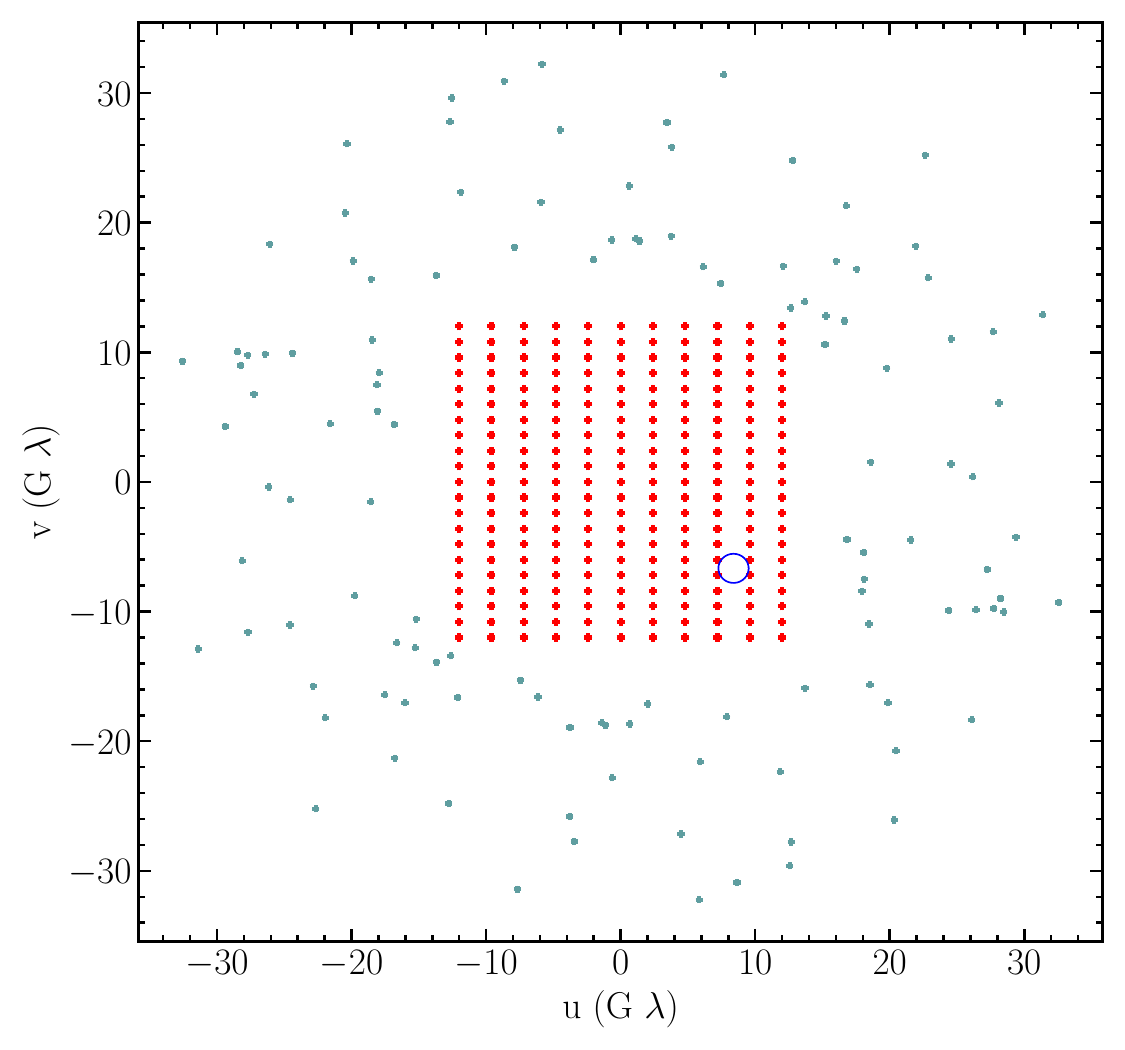}
\includegraphics[width=0.32\textwidth]{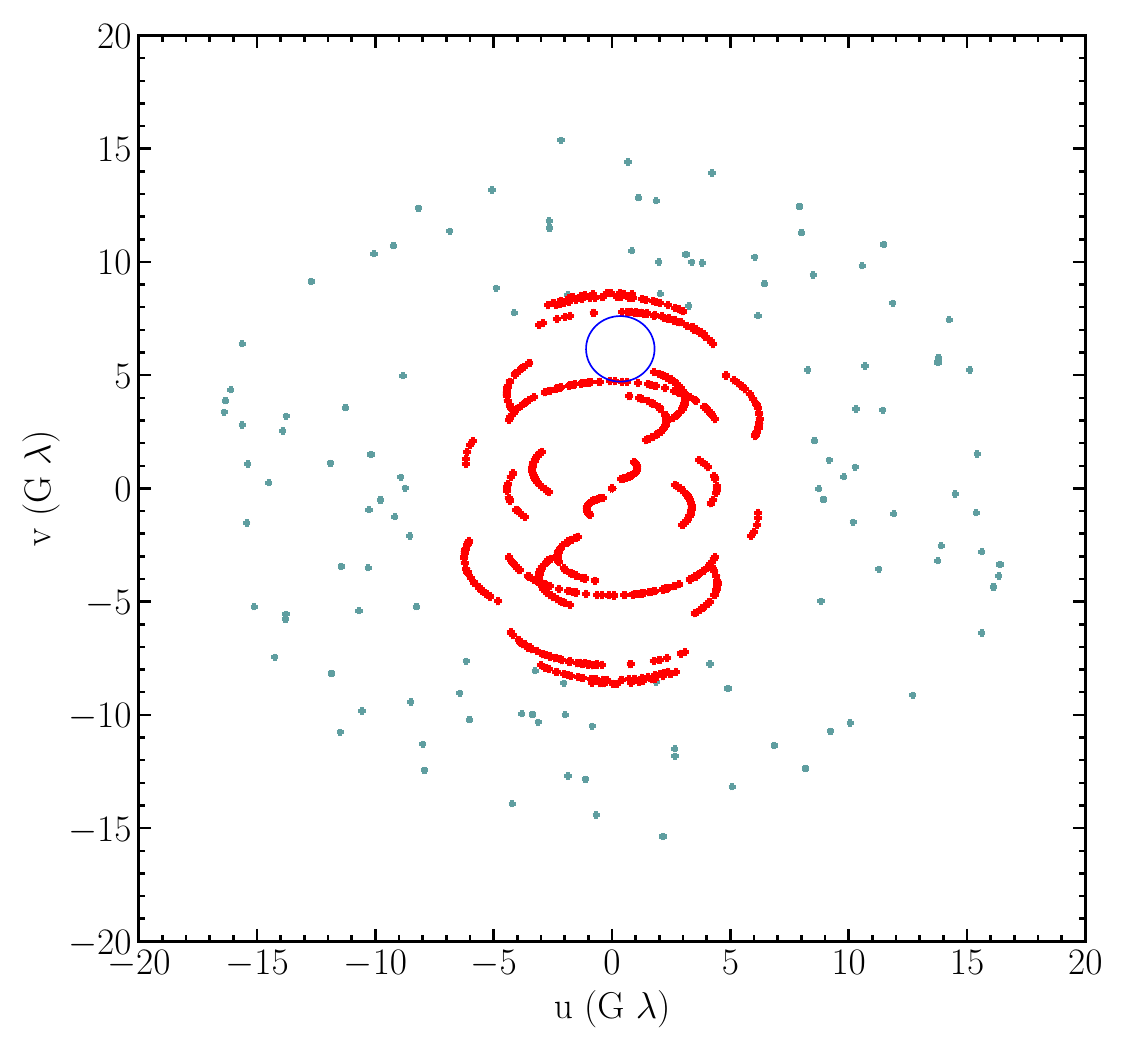}
\includegraphics[width=0.32\textwidth]{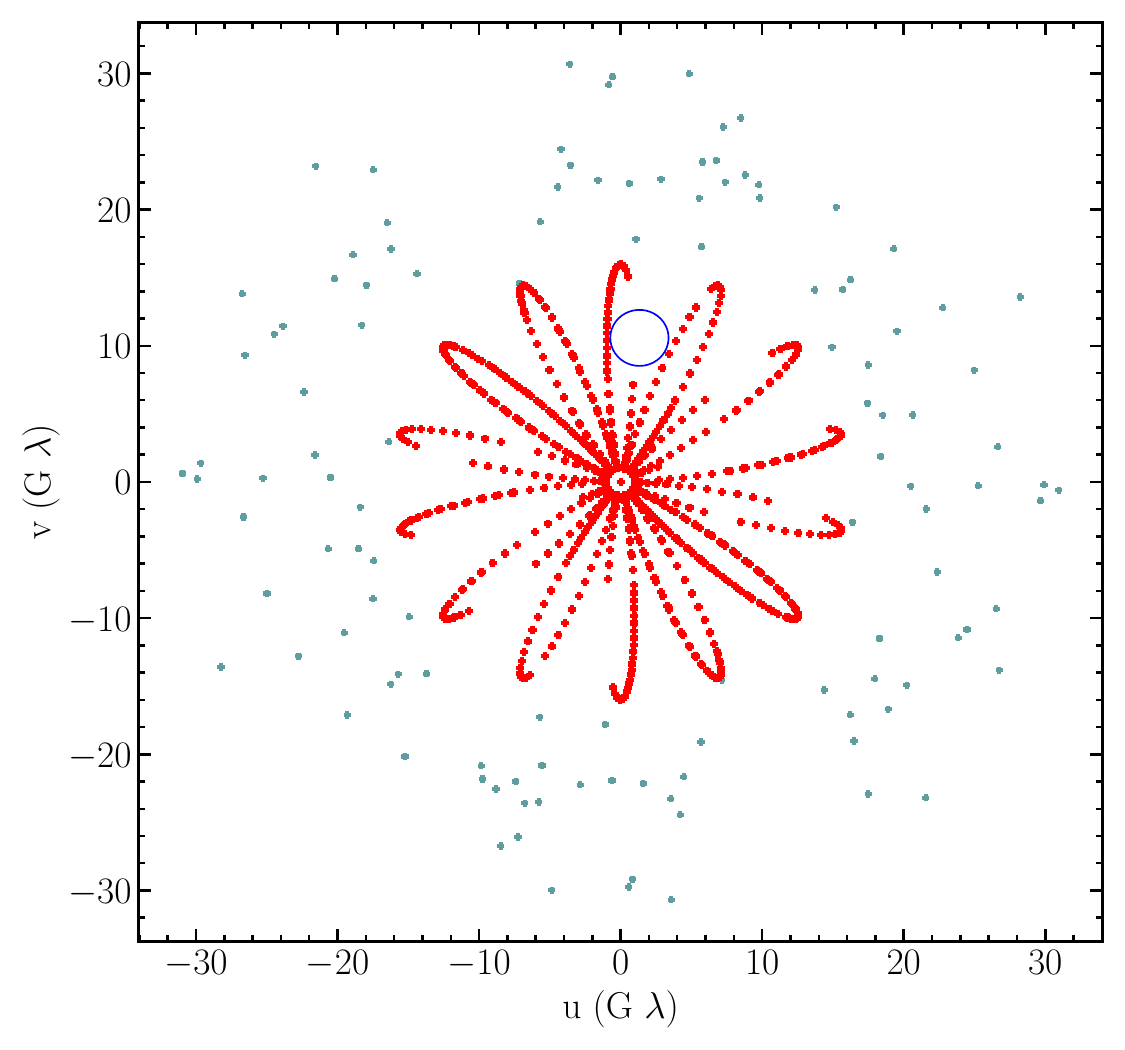}
\caption{\footnotesize Three examples of sparse interferometric coverage: (left) uniformly spaced, (middle) EHT array for 2017 observations of Sgr~A$^*$, (right) simple representation of two satellites at different orbits. In each panel, the red points represent the locations of the measurements, whereas the blue points show the randomly located ghost points used to suppress the high spatial frequencies. The blue circles identify the size of a typical gap in the $u-v$ coverage.}
\label{fig:patterns}
\end{figure*}

\subsection{Kernel Selection}

Kernels can be constructed out of numerous functions, such as polynomials, Gaussians, or sigmoids, using the definition in equation~\ref{eq:kernel}. Alternatively, one can write a functional form for the kernel itself, as long as the corresponding Gram matrix is positive semidefinite, i.e.,
$V^T K V \ge 0$ for any vector $V$ in $R^n$. 

Mercer’s theorem states that if $\lambda_i$ and $\phi_i$ are the eigenvalues and eigenfunctions of $k(x,x^\prime)$, then 
\begin{equation}
k(x,x^\prime)=\sum^K_{i=1} \lambda_i \phi_i (x) \phi_i^*(x^\prime)
\end{equation}
Because of this, using a kernel method is equivalent to interpolating between data points using a function that is a linear combination of the eigenfunctions of the kernel. Kernels can have an infinite number of eigenvalues and corresponding eigenvectors, in which case they are called non-degenerate kernels. This approach, therefore, has the significant advantage that, even though the kernel may have only a few hyperparameters that need to be inferred from the data, the interpolation is equivalent to utilizing an infinite number of functions. The small number of hyperparameters also obviates the need for regularizers in the regression. 

In this paper, we use the Matern kernel
\begin{equation}
    K(d)=\frac{2^{1-\nu}}{\Gamma(\nu)}
    \left(\sqrt{2\nu}\frac{d}{\rho}\right)^\nu
    K_\nu\left(\sqrt{2\nu}\frac{d}{\rho}\right)\;,
    \label{eq:Matern}
\end{equation}
where $d$ is the Euclidean distance between the two points, $\Gamma(\nu)$ is the $\Gamma$ function,  $K_\nu$ is the modified Bessel function of the second kind, and $\nu$ and $\rho$ are the two hyperparameters. This kernel is non-degenerate and is invariant to translations (i.e., stationary). The Matern kernel reduces to the exponential kernel for $\nu$ = 0.5 and to the Gaussian kernel for $\nu\rightarrow \infty$. Therefore, the Matern kernel includes a large class of kernels and is very useful for many applications because of this flexibility.

\section{The \krisp\ algorithm}

Interferometric data consists of complex visibilities measured at different baseline locations. In this initial presentation of the algorithm, we assume that the real and imaginary components of the visibilities can be individually measured. The measurement of complex phases is currently a challenge for ground-based interferometers at millimeter or optical wavelengths due to atmospheric turbulence, which instead rely on the sum of phases along closed baseline triangles referred to as closure phases~\citep{Kulkarni1989,Rogers1995}. 

Even though we can, in principle, perform a single kernel regression for complex visibilities, in practice, the real and imaginary parts of the visibilities can have different correlation lengths. For example, the correlation length in the real components of the visibilities for a compact symmetric source is inversely proportional to the size of the image, while the correlation length in the imaginary components is practically infinite (as they are all zero). For this reason, we perform Kernel regression separately to the real and imaginary parts of the visibilities. 

A second advantage of performing the regression to the real and imaginary components of the visibilities, as opposed to the amplitudes and phases, is that the measurements of the individual components have Gaussian uncertainties arising from thermal noise. This property allows us to employ methods of Gaussian Random Processes both to incorporate the measurement uncertanties in the regression and also to quantify the uncertainty in the reconstructed Fourier maps (see eqs.~[\ref{eq:Vpred}]-[\ref{eq:Verror}]). 

In this implementation, we assume that data points have uncorrelated Gaussian errors with the same variance. Real data can be both heteroscedastic, because the telescopes in an array are typically not identical, and correlated, because of residual errors in the gain calibrations of individual telescopes. Kernel methods can handle such more complex situations \citep{Cawley2004}, which we will explore in future work.  

We implement the kernel regression method described in equations~(\ref{eq:Vpred}) and (\ref{eq:Verror}) using the Matern kernel in equation~(\ref{eq:Matern}) and Gaussian uncertainties for the data points. We fix one of the parameters of the Matern kernel to $\nu=1.5$ and use the \texttt{GaussianProcessRegressor} function in the \texttt{scikit-learn} library to optimize the hyperparameter $\rho$ of the kernel based on the data, which follows Algorithm~2.1 of \citet{Rasmussen2006}. 

A last important consideration in the interferometric reconstruction algorithm is how to handle the lack of information at baseline lengths beyond the largest baseline length $b_{\rm max}$ probed by the array. Different low pass-band filters have been used to suppress the Fourier components at beyond the largest baselines, each of which have different advantages and disadvantages. For example, the traditional Gaussian filter decays gradually to zero but also suppresses power substantially within the domain of observation. A Butterworth filter, on the other hand, preserves most of the observed information but may introduce small-scale ringing in the reconstructed image \citep{Psaltis2020}. 

Kernel methods provide an elegant alternative way to reduce the influence of baseline lengths for which there is no observational coverage. In the \krisp\ algorithm, we introduce a number of randomly located (in baseline length and orientation) ``ghost'' data points at baseline lengths between $b_{\rm max}$ and $2\; b_{\rm max}$ at which we set the value of the complex visibilities to zero. This naturally removes the high spatial-frequency content from reconstructed images while ensuring a gradual transition to zero beyond $b_{\rm max}$. For every ghost point we also introduce its conjugate, to ensure that the reconstructed Fourier map will have the conjugate symmetry that corresponds to a real-valued image.

\section{Test Configurations and Images}

While the \krisp\ algorithm provides a general tool that can be used for many sparse interferometric reconstructions, we choose as our examples horizon-scale black hole images and the very large array configurations required to obtain such images, as these provide some of the most challenging cases for sparse imaging. 

\begin{figure*}[t]
\centering
\includegraphics[width=0.65\textwidth]{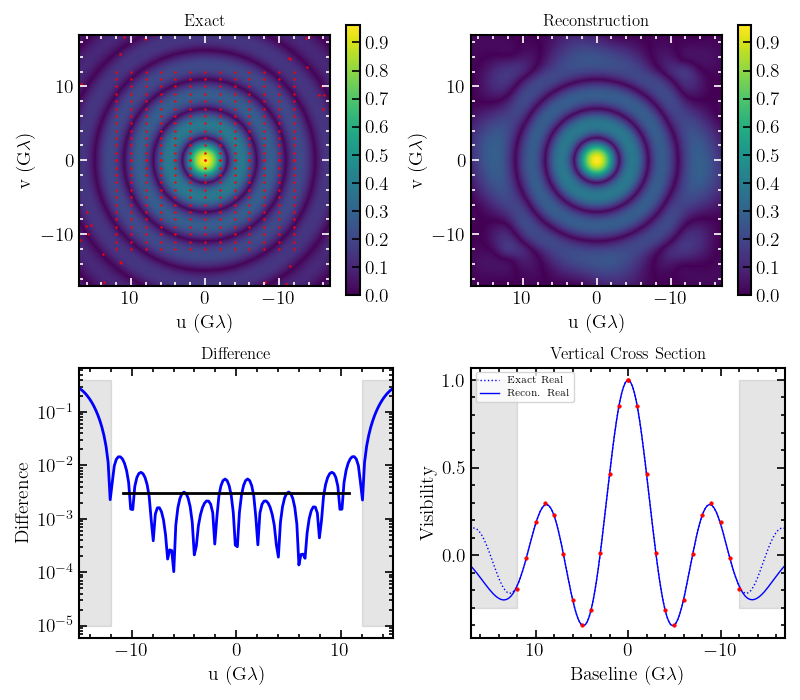}
\caption{\footnotesize (Top left) The 2D Fourier transform of an infinitesimally thin ring with and an example of a uniform 2D grid used for sampling the function. (Top right) The approximate function reconstructed using Kernel methods throughout the observed domain. (Bottom left) The difference between the exact and the approximate function. The black horizontal line shows the rms error calculated on the observed domain. (Bottom right) The horizontal cross section of the exact and approximate functions. The approximate function starts to deviate from the exact one in the region without any data points, shown as the vertical grey bands in the bottom panels. The red points denote the locations of the sampling.}
\label{fig:uniform_grid_points}
\end{figure*}

\begin{figure}[t]
\centering
\includegraphics[width=0.9\columnwidth]{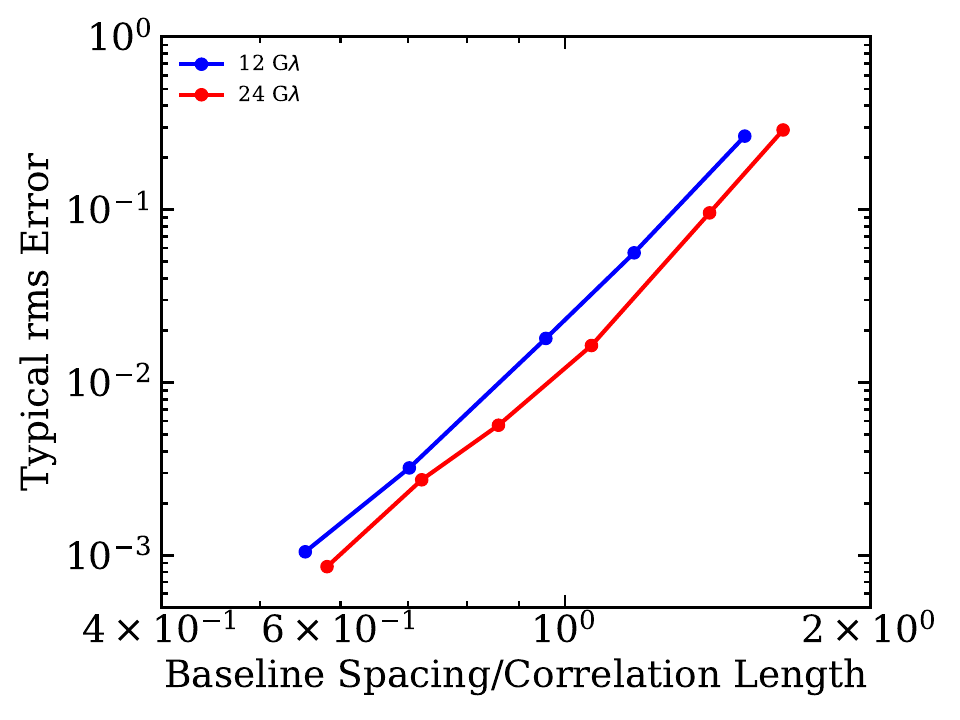}
\caption{\footnotesize The typical rms error between the Fourier map of an infinitesimally thin ring and its reconstruction obtained from sampling on a uniform grid plotted against the spacing between successive baselines divided by the correlation length of each Fourier map. The blue and red curve correspond to reconstructions with maximum baseline length of 12~G$\lambda$ and 24~G$\lambda$, respectively. The error remains $\lesssim$1\% for baseline spacings smaller than the correlation lengths of the Fourier maps.}
\label{fig:uniform_error}
\end{figure}

\subsection{Examples of Sparse Interferometric Coverage} 

We introduce three distinct configurations of sparse interferometric coverage that will allow us to evaluate the performance of our algorithm (see Fig.\ref{fig:patterns}). 

The first is a pattern that is uniformly spaced in both E-W and N-S directions. Even though this is not motivated by the $u-v$ plane coverage of a realistic interferometer, it generates regular gaps and allows us to explore the impact of the gap size on the image reconstruction. 

The second is the $u-v$ coverage of the Earth-based EHT array for the 2017 observations of Sgr A$^*$, the black hole at the center of the Milky Way. A difference with respect to the real EHT observations is the assumption that complex visibilities, and not just closure phases, can be measured. We will explore extension to closure phases in future work.  

The last configuration is motivated by the $u-v$ plane coverage that can be obtained from two satellites orbiting at different altitudes and inclinations~\citep{BenZineb2024}. This configuration leads to repetitive shapes with large excursions in baseline length and small azimuthal drift caused by the relative difference in orbital speeds, which helps cover the $u-v$ plane. To keep the formulation as general as possible for the analyses in this paper, we represent the u-v coverage of such a two-satellite configuration by utilizing an analytic function of the form  
\begin{eqnarray}
    (u,v)&=&(b\cos\theta,b\sin\theta)\nonumber\\
    b&=&1+\epsilon \cos(n\theta),\label{eq:rose}
\end{eqnarray}
which captures the shapes of typical tracks and is flexible for representing different types of orbits by changing the parameters $\epsilon$ and $n$. In addition, we introduce randomly placed gaps that lead to partial coverage in the $(u,v)$-tracks to mimic the effects of realistic scheduling of observations.

For each of these configurations, we quantify the size of typical gaps using the following measure. At each point in the $(u,v)$ plane with a baseline length that is less than the maximum allowed by the particular configuration, we prescribe the largest circle that does not intersect any data points. We then calculate the 75th percentile radius of all the circles weighted by their surface areas and use this as a measure of the typical gap size, which we denote in Figure~\ref{fig:patterns} by a blue circle.

\subsection{Geometric Models for Black Hole Images}

To generate simple synthetic data that capture the main characteristics of horizon-scale black hole images, we use the analytic model of \citet{Kamruddin2013}. In this geometric construction, we use two offset disks of different radii $R_1$ and $R_2$ to create crescent images. We define the mean diameter of the ring as $D\equiv 0.5 (R_1+R_2)$, its relative thickness as $\psi \equiv 1- (R_1/R_2)$, and the degree of asymmetry (for a crescent) as $\tau\equiv 1- (a_0^2+b_0^2)^{1/2}/(R_1-R_2)$. Here, $(a_0,b_0)$ is the offset of the center of the inner ring from the origin, which introduces an orientation of brightness asymmetry at an angle $\phi=\tan^{-1}(b_0/a_0)$ measured East-of-North. Note that, when $(a_0,b_0) = (0,0)$, the image becomes a symmetric ring, while in the limit of $\psi \rightarrow 0$, the ring attains infinitesimal thickness. 

The normalized complex visibility of the general crescent can be constructed out of the Fourier transforms of the two disks and is given by
\begin{equation}\label{eq:crescentvisibility}
V(u,v)=\frac{R_1 J_1(k R_1)}{k}-e^{-2\pi i (a_0 u+b_0 v)}\frac{R_2 J_1(k R_2)}{k}\;,
\end{equation}
where $u$ and $v$ are the spatial frequencies measured in units of the wavelength $\lambda$ of observation, $k \equiv \sqrt{u^2 + v^2}$, and $J_1$ is the first Bessel function of the first kind.

\section{Impact of Sparseness on Fourier Domain Reconstruction with Kernel Methods}

In this section, we will explore the effect of the sampling density in $u-v$ space on the reconstruction of the two-dimensional Fourier transform of an image. To illustrate the principles, we will use an infinitesimally thin ring as the underlying image and use the test configurations introduced in the previous section. 

Figure~\ref{fig:uniform_grid_points} shows the case of a uniform grid. The four panels depict the Fourier transform of the exact image and the locations of the data points (top left), the Fourier domain map recovered using \krisp\ (top right), a cross section of the difference between the two (bottom left), and the cross section of the ground truth and reconstructed Fourier transforms (bottom right). In this example, the 2~G$\lambda$ spacing between the data points is slightly smaller than the correlation length of the Fourier transform of the ground-truth image. The latter can be approximately inferred from the baseline length at which the visibility amplitude drops to half of its maximum value, which is $\sim2.2\;$G$\lambda$, as can be seen in the cross section shown in Figure~\ref{fig:uniform_grid_points}. 

\begin{figure*}[t]
\centering
\includegraphics[width=0.65\textwidth]{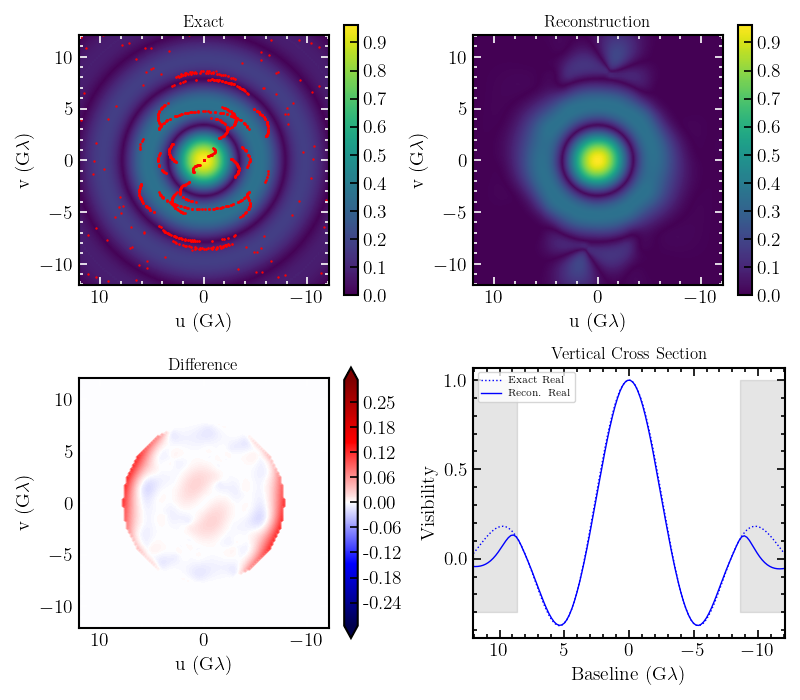}
\caption{\footnotesize Same panels as Fig.~\ref{fig:uniform_grid_points} but for the $u-v$ plane coverage for Sgr~A$^*$ by the 2017 EHT array. The bottom left panel shows the difference between the exact and approximate Fourier maps up to the maximum baseline length for which data points exist.}
\label{fig:EHT_points}
\end{figure*}

The difference between the ground truth and reconstructed Fourier maps is very small throughout the region where data points exist. In fact, it deeps to negligible values at the locations of the data points and rises in between them, as expected. Because the distance between successive data points is smaller than a correlation length, the frequent anchoring at the data points prevents the difference to grow significantly. To help quantify the degree of fidelity of the reconstruction, we define the ``typical'' error as the rms difference between the ground truth and the reconstruction, measured in the domain up to a fraction (typically 0.9) of the maximum baseline length at which data points exist. We show this as the horizontal black line in the bottom left panel of Figure~\ref{fig:uniform_grid_points}. 

To demonstrate the performance of \krisp, we repeat the reconstruction of the thin ring with uniform $u-v$ coverage of varying separation and show the typical rms error in Figure~\ref{fig:uniform_error}. The error remains below $\sim 1\%$ for baseline spacings that are smaller than the correlation length of the Fourier maps. We also show that varying the maximum baseline length in the E-W and N-S orientations from 12$\;$G$\lambda$ to 24$\;$G$\lambda$ makes a minor difference in the rms error. 

Next, we use a thin ring ground truth image with the $u-v$ coverage of the 2017 EHT array for Sgr~A$^*$. Figure~\ref{fig:EHT_points} shows the same four panels as Figure~\ref{fig:uniform_error} except for the lower left panel, which is a 2-D map of the difference between the ground truth and the \krisp\ reconstruction. Despite the significant differences in the locations (or tracks) of the data points in Fourier space between the two configurations, the error in the reconstructed map remains below $\sim 7\%$ in the regions where the data points exist.

\begin{figure*}[t]
\centering
\includegraphics[width=0.7\textwidth]{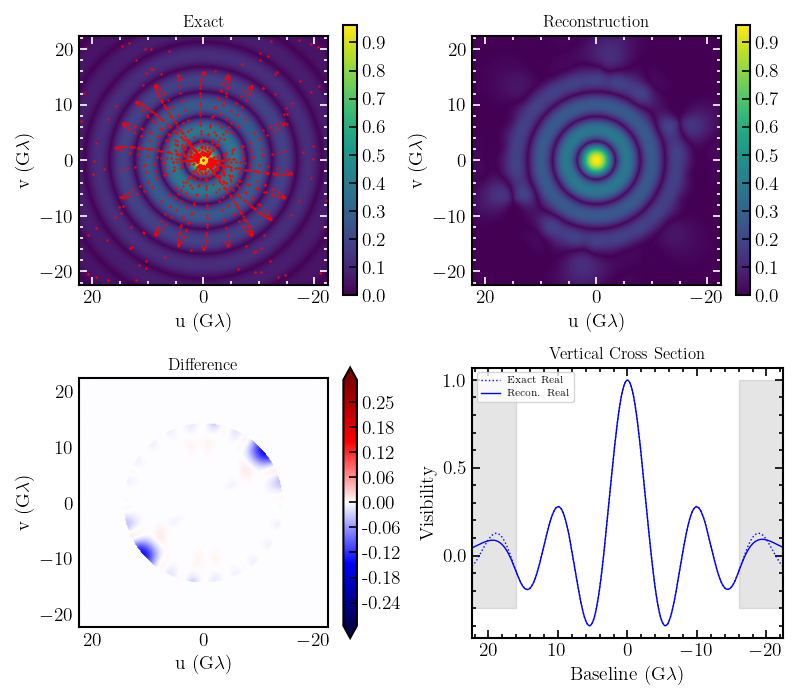}
f5\caption{\footnotesize Same panels as Fig.~\ref{fig:EHT_points} but for tracks on the u-v plane that mimic an interferometer consisting of two satellites.}
\label{fig:rose_points}
\end{figure*}

Finally, we show in Figure~\ref{fig:rose_points} the performance of \krisp\ for the Fourier domain tracks resembling a partial coverage from two satellites on different orbits. The error at the location of the tracks is again negligible and rises marginally between them, leading to a high-fidelity reconstruction. The only region where the difference rises to an appreciable value, $\sim 10\%$, arises from a gap that simulates a lack of coverage that could result from, e.g., source visibility or satellite operations. 

\begin{figure}[t]
\centering
\includegraphics[width=0.9\columnwidth]{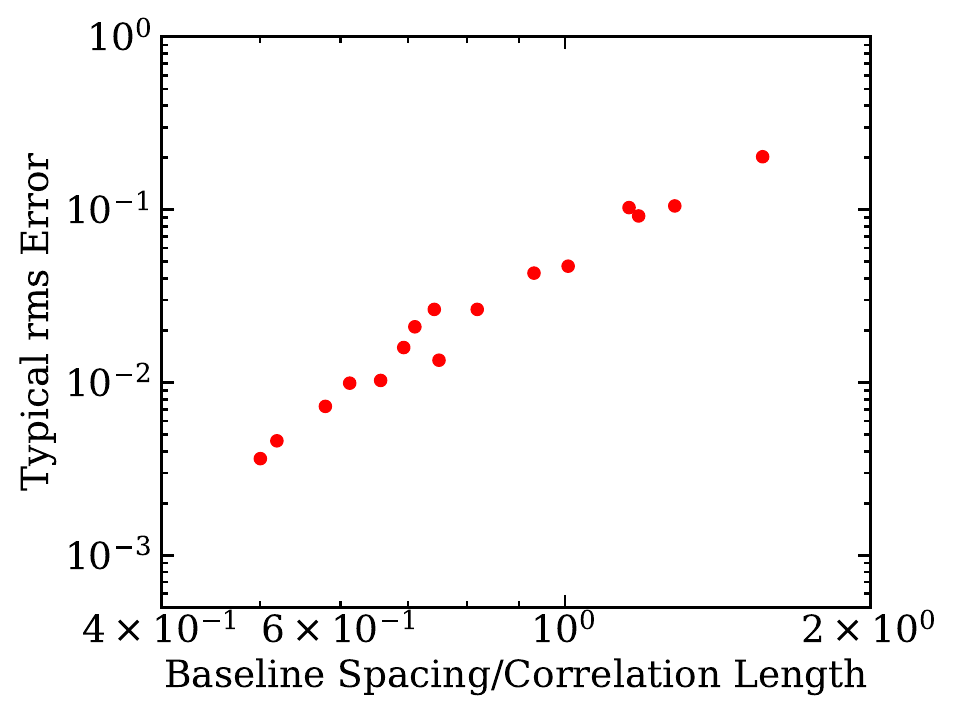}
\caption{\footnotesize Same as Fig.~\ref{fig:uniform_error} but with the sampling of the $u-v$ plane with the two-satellite configuration. The typical spacing of the data points shown on the x-axis is varied by changing the pattern of the tracks as well as the partial coverage of the tracks. The rms error in the reconstructed function remains below $\approx 5\%$ as long as the gaps in the $u-v$ plane remain below one correlation length.}
\label{fig:rose_error}
\end{figure}

We repeated the reconstructions with this last configuration but changing the parameter $n$ that controls the number of leaves in the $u-v$ coverage pattern (see eq.~[\ref{eq:rose}]) and the number of data points along each segment such that the typical size of the data gaps changes. Figure~\ref{fig:rose_error} shows the dependence of the typical rms error we defined earlier plotted against the ratio of the typical baseline spacing to the correlation length of the image. As in the case of the uniform spacing, the fidelity of the \krisp\ reconstruction depends strongly on this ratio, with the error reaching $\sim 1\%$ for baseline spacings that are equal to $\approx 0.5$ correlation length for the image.  

In both Figures~\ref{fig:EHT_points} and \ref{fig:rose_points}, the Kernel method reconstruction starts to deviate significantly from the ground-truth function beyond the domain where data exist. The use of ghost points in the domain of extrapolation helps suppress these deviations. More importantly, in the next section where we introduce the effect of measurement errors and use them to predict the uncertainties in the reconstruction, it will become evident that the method identifies naturally the regions of large potential deviations, without the need to know the ground truth a priori. 

\section{Image Reconstruction with Measurement Errors and Complex Images}

In the previous section, we focused on the impact of multiple types of sparse $u-v$ coverage and showed that the reconstruction of the Fourier maps with \krisp\ depends primarily on the ratio of the typical baseline spacing to the correlation length in Fourier space, where the latter is directly determined by the size of the target image. In this section, we turn to exploring the influence of measurement errors and more complex image structures on the fidelity of image reconstruction. To that end, we first introduce images that do not possess azimuthal symmetry, for which the Fourier transforms have both real and imaginary components. These images also have substantial azimuthal structure in their Fourier maps and allow us to explore the performance of the method in such cases. 

\begin{figure*}[t]
\centering
\includegraphics[width=0.7\textwidth]{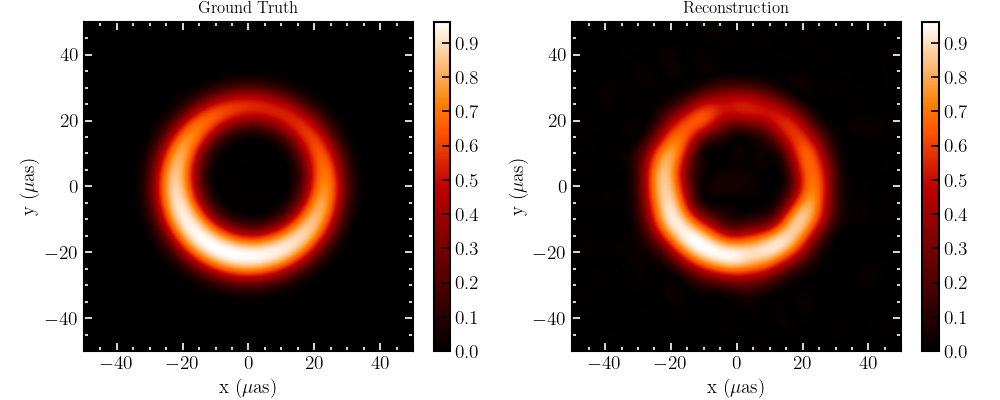}
\caption{\footnotesize The ground truth and \krisp\ reconstruction for an example asymmetric crescent image with data uncertainties and Fourier maps shown in Fig.~\ref{fig:realistic_6panel}.}
\label{fig:realistic_images}
\end{figure*}

\begin{figure*}[t]
\centering
\includegraphics[width=0.95\textwidth]{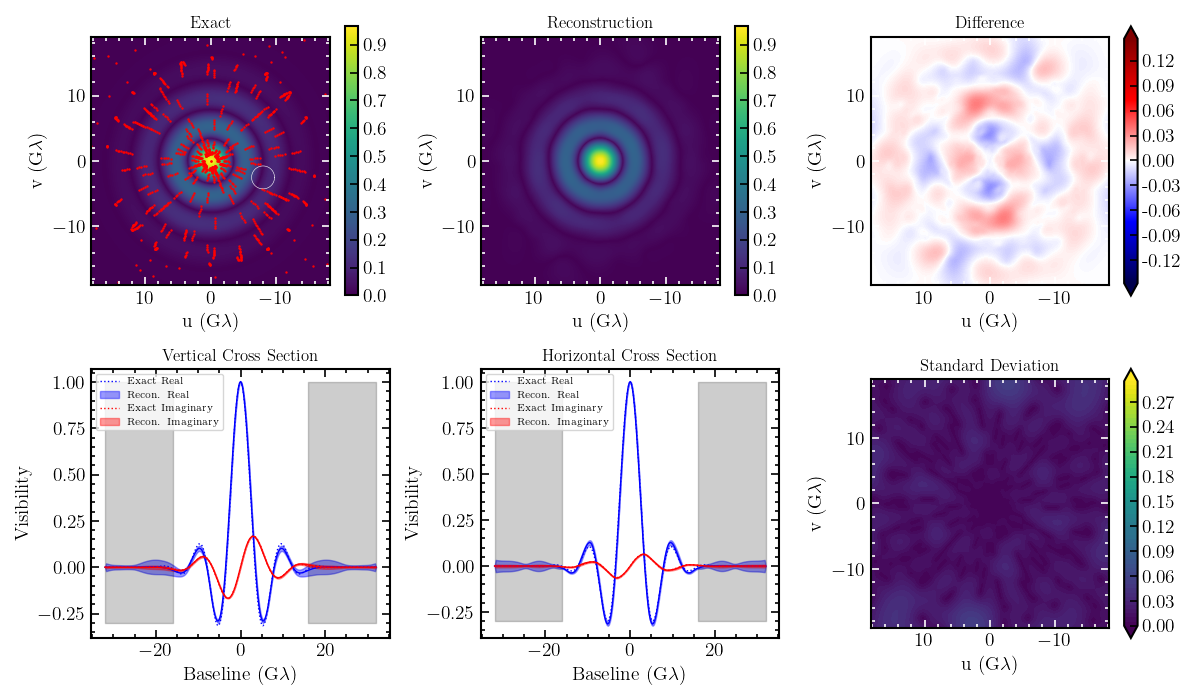}
\caption{\footnotesize The three panels in the top row show the ground-truth and reconstructed Fourier maps as well as the difference between the two for an asymmetric crescent image described in the text and shown in Fig.~\ref{fig:realistic_images}. The panels in the bottom row show the vertical and horizontal cross sections of the Fourier maps and the standard deviation in the predicted reconstruction that arises from the formal uncertainties in the data. The bands in the real and imaginary parts of the reconstructed cross sections depict one standard deviation.}
\label{fig:realistic_6panel}
\end{figure*}

\subsection{Impact of Measurement Errors}

As a first example of reconstruction with measurement errors, we use a ground truth image of a crescent with a diameter of $D=46~\mu$as, relative width of $\psi=0.3$, brightness asymmetry of $\tau=0.6$, oriented at an angle $\phi=20^\circ$ EofN, and broadened with a Gaussian of $\sigma_{\rm G}=8~\mu$as (see left panel of Fig.\ref{fig:realistic_images}). These parameters are comparable to the values inferred for the 1.3$\;$mm image of the black hole at the center of the M87 galaxy~\citep{EHT_M87_1}. The correlation length of this image is approximately 2.2$\; {\rm G}\lambda$ in all orientations.

\begin{figure*}[t]
\centering
\includegraphics[width=0.95\textwidth]{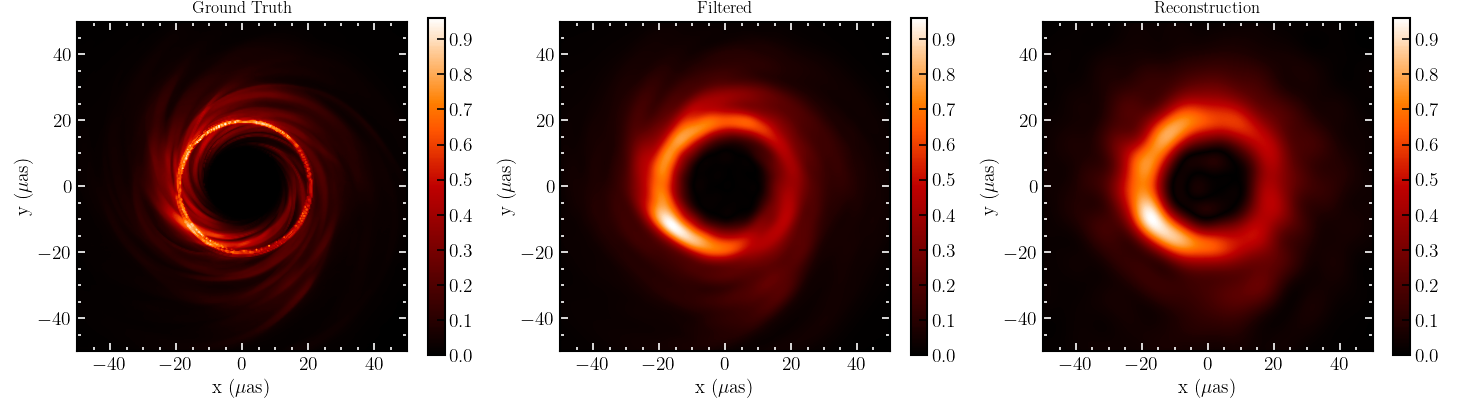}
\caption{\footnotesize (Left) Example image from a snapshot of a GRMHD simulation; (Middle) the same image filtered at the resolution of the maximum baseline of an array with sparse sampling shown in Fig.~\ref{fig:ex64_6panel}; (Right) \krisp\ reconstruction of the image that recovers the asymmetry and substructure.}
\label{fig:ex64_images}
\end{figure*}

\begin{figure*}[t]
\centering
\includegraphics[width=0.95\textwidth]{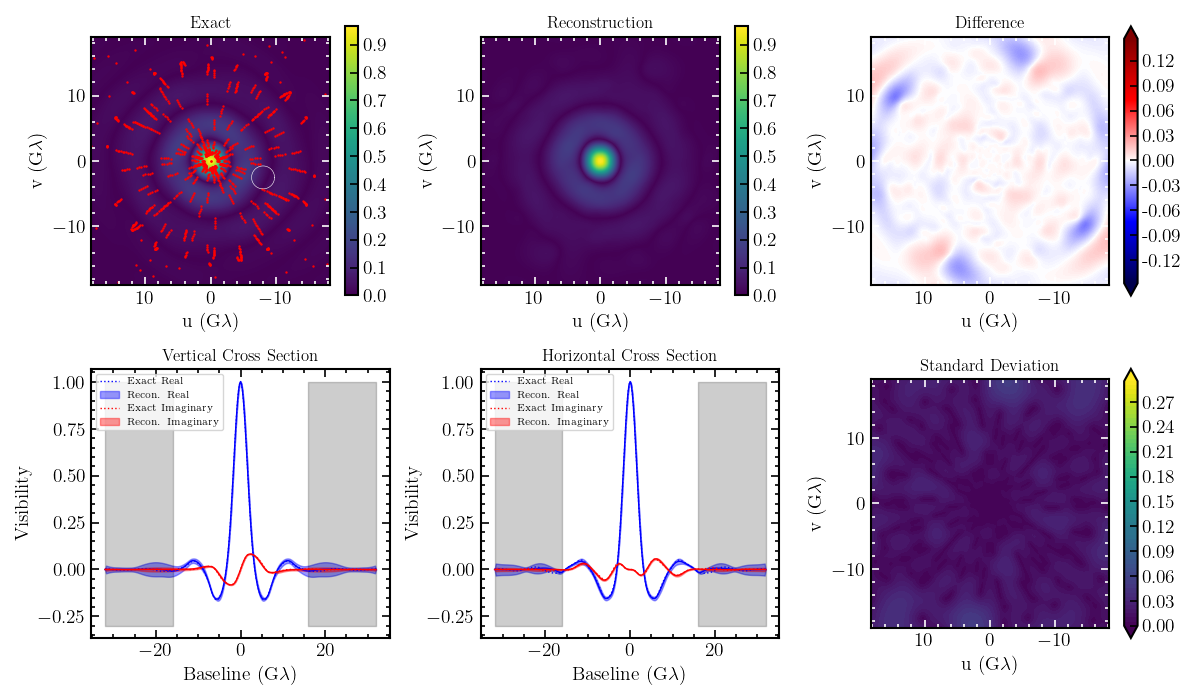}
\caption{\footnotesize Same as Fig,~\ref{fig:realistic_6panel} for the example shown in Fig.~\ref{fig:ex64_images}.}
\label{fig:ex64_6panel}
\end{figure*}

We choose a configuration of sparse $u-v$ coverage with the rose pattern that reaches a maximum baseline length of 16$\;$G$\lambda$ and has a typical baseline spacing of 0.8 times the correlation length of the image. We introduce Gaussian errors to all baselines with standard deviations equal to 1\% of the peak visibility amplitude. This corresponds to an SNR of 100 at the shortest baselines and SNR$\simeq 3$ around a baseline length of 14$\;$G$\lambda$, which is near the maximum baseline.

We show in Figure~\ref{fig:realistic_6panel} the ground-truth and reconstructed Fourier maps, the difference between the two, the vertical and horizontal cross sections of the Fourier maps, and the standard deviation in the predicted reconstruction. This last quantity is calculated using equation~(\ref{eq:Verror}), with the $\sigma^2$ set equal to the formal variance in the measurement. The error bands displayed in the two cross section panels correspond to one standard deviation in the reconstructed map.

The standard deviation map shows that the uncertainty in the prediction of the Fourier map reconstruction is affected both by the formal errors and by the relative spacing of the data points. This uncertainty is equal to the formal error at the locations of the data points, as expected, and increases by a factor of few between them, as can be seen in the lower right panel of Figure~\ref{fig:realistic_6panel}. For this reason, the width of the bands shown in the two cross sections increases with baseline length, both because the SNR of the data points there is smaller and because the spacing between data points increases with baseline length in this configuration. 

\begin{figure*}[t]
\centering
\includegraphics[width=0.95\textwidth]{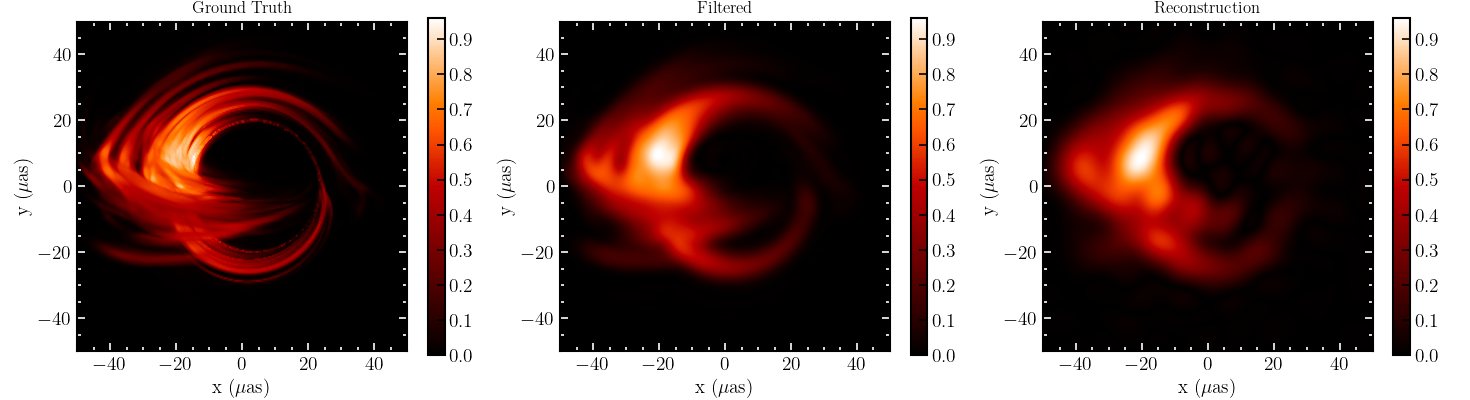}
\caption{\footnotesize Same as Fig.~\ref{fig:ex64_images} for a different image from a GRMHD simulation that has substantially different thickness, asymmetry, and substructure.}
\label{fig:ex8_images}
\end{figure*}

\begin{figure*}[t]
\centering
\includegraphics[width=0.95\textwidth]{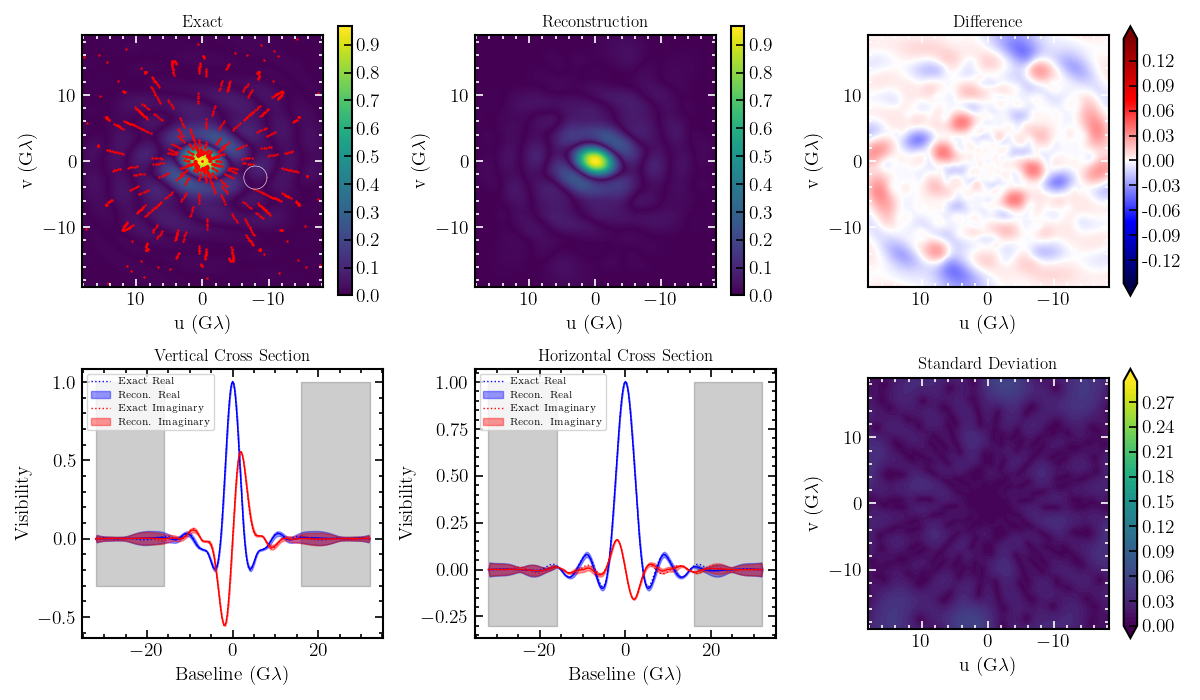}
\caption{\footnotesize Same as Fig.~\ref{fig:ex64_6panel} for the image shown in Fig.~\ref{fig:ex8_images}.}
\label{fig:ex8_6panel}
\end{figure*}

We also show in Figure~\ref{fig:realistic_images} the ground-truth image (left panel) and the reconstructed image in the presence of Gaussian errors. The Kernel method yields a very accurate reconstruction of the Fourier map and the image, with a maximum difference in the former of a few percent throughout the observational domain. 

\subsection{GRMHD Images}

We consider next images that possess complex radial and azimuthal structures, as expected for black hole images owing to the magnetic field structure and turbulent motions of the plasma in the inner accretion flow. General Relativistic Magnetohydrodynamic simulations have been used to explore the characteristics of such images with a higher degree of complexity, which we use as synthetic data in this section (see, e.g., the library in \citealt{EHT_SgrA_V}). 

In the example we show in Figure~\ref{fig:ex64_images}, we use the 1.3 mm image from a snapshot of a simulation with a black hole spin $a=0.5$, MAD magnetic field configuration, electron temperature parameter $R_{\rm high} =40$, and observer's inclination of 30$^\circ$ (see \citet{EHT_SgrA_V}). We introduce, as before, Gaussian errors with a standard deviation of 0.003 of the peak visibility amplitude such that the average SNR in the largest baselines is 3. We sample the $u-v$ space up to 16$\;$G$\lambda$ with a rose pattern that we show in the top left panel of Figure~\ref{fig:ex64_6panel}, which has a ratio of typical baseline spacing to the correlation length of the image of $\sim 1$.

We use \krisp\ to reconstruct the Fourier maps and the resulting images and show the results in the same figures. The reconstruction error in the Fourier domain is at most a few percent. Consequently, the resulting image characteristics have high fidelity in reproducing the properties of the ground truth. These include the size and the width of the image, the overall brightness asymmetry, and substructure at scales probed by the larger baselines of the array. 

Finally, we explore an even more complex image structure that we show in Figure~\ref{fig:ex8_images}. This snapshot is from a GRMHD simulation with a black hole spin of $a=0.5$, a SANE magnetic field configuration, electron temperature parameter $R_{\rm high}=10$ and an observer's inclination of $70^\circ$. As can be seen in the figure, it has substantial brightness asymmetry as well as significantly different substructure and thickness from the previous example. We keep the identical $u-v$ coverage and measurement errors as in the previous example (see Fig.~\ref{fig:ex8_6panel}).   

The brightness asymmetry in the image introduces significant difference between the correlation lengths in the Fourier space along the two principal directions of the image, which, in this case, are aligned with E-W and N-S orientations. This is visible in the top middle panel of Figure~\ref{fig:ex8_6panel}. As a result, the ratio of the mean baseline spacing to the correlation length in this example varies between 0.7 in the E-W direction and $\sim 1$ in the N-S direction. Even though, in this implementation, \krisp\ utilized only a single correlation length in all orientations, the resulting reconstruction of both the Fourier map and of the resulting image reproduces with very high fidelity all the characteristics of the ground truth up to the resolution of the array.

\section{Conclusions}

We introduced \krisp, a new kernel-method based imaging algorithm for sparse interferometric arrays. We focused on the reconstruction of horizon-scale black hole images as the combination of the sparseness of the required arrays and the presence of significant image structure provide challenging test cases. We showed that \krisp\ generates high fidelity reconstructions as long as the typical gaps in the interferometric coverage do not exceed the correlation length of the Fourier map of the underlying image.  

In this first implementation of the algorithm, we assumed that the measurement uncertainties are the same for all baselines that interferometric phases can be directly measured, which is currently possible in space based or long-wavelength terrestrial interferometers. However, there is no intrinsic limitation to the method for incorporating heteroscedastic and/or correlated errors (e.g., to account for gain uncertainties), which we will explore in forthcoming work. 

\bibliographystyle{aasjournal}

\bibliography{kernel_methods}{}

\end{document}